\documentclass[11pt]{article}
\UseRawInputEncoding
\usepackage{verbatim}
\usepackage[colorlinks=false]{hyperref} 
\usepackage{subfigure}
\usepackage{pgfplots}
\usepackage{verbatim}
\usepackage{epstopdf}

\usepackage{amsmath}
\allowdisplaybreaks[4]

\newcommand{\eq}{\begin{equation}}
\newcommand{\en}{\end{equation}}
\newcommand{\eqa}{\begin{eqnarray}}
\newcommand{\ena}{\end{eqnarray}}
\newcommand{\eqs}{\begin{displaymath}}
\newcommand{\ens}{\end{displaymath}}
\newcommand{\eqas}{\begin{eqnarray*}}
\newcommand{\enas}{\end{eqnarray*}}

\usepackage{graphicx}
\usepackage{amsmath}
\usepackage{mathrsfs}
\usepackage{amssymb}
\usepackage{bm}
\usepackage{titlesec}
\usepackage{array}
\usepackage{enumerate}
\usepackage{tikz}
\usepackage{CJK}
\usepackage{epsfig}
\usepackage{amsfonts,amsthm}
\usepackage{pdfpages}

\usetikzlibrary{shapes,snakes}

\usepackage{multirow, makecell}
\usepackage{appendix}

\numberwithin{equation}{section}
\titleformat{\section}[hang]{\large\bfseries}{\\ \thesection\,}{0.8em}{}{}
\titleformat{\subsection}[hang]{\normalsize\bfseries}{\\ \thesubsection\,)}{0.8em}{}{}
\titleformat{\subsubsection}[hang]{\bfseries}{\\ \thesubsubsection\,)}{0.8em}{}{}

\advance\oddsidemargin by -\textwidth
\advance\oddsidemargin by -1in
\evensidemargin=\oddsidemargin

\begin{document}

\begin{center}

   {\bf \large
    Quantum Computation Using Action Variables
 }
 \vspace{.5cm}

{\small Yong Zhang \footnote{yong\_zhang@whu.edu.cn. }
and Konglong Wu
\footnote{Under the guidance of the first author, the second author had performed a Bachelor thesis study and a graduate study on the topics of the paper
 from 2016 to 2018. }
 }\\[3mm]

School of Physics and Technology,  Wuhan University, Wuhan 430072, P. R. China  \\[1cm]

\end{center}

\begin{center}

 { \bf Abstract }\\[3mm]

\parbox{13.cm}{Recently, Lloyd and Montangero have made a brief research proposal on universal quantum computation
in integrable systems. The main idea is to encode qubits into quantum action variables and build up quantum gates
by the method of resonant control. We study this proposal to argue quantum computation using action variables as
fault-tolerant quantum computation, whose fault-tolerance is guaranteed by the quantum KAM theorem. Besides, we view
the Birkhoff norm form  as  a mathematical framework of the extended harmonic oscillator quantum computation.
\\

{\bf Key Words}

Quantum Computing, Integrable system, Action variable, KAM theorem, Birkhoff normal form \\

 {\bf PACS numbers}   03.65.Ud, 02.10.Kn, 03.67.Lx}

\end{center}



\newpage

\section{Introduction}

Quantum information and computation has induced a modern development of quantum mechanics.
As a physical process in a quantum mechanical system, quantum computing has the computational power
beyond classical computing due to quantum superposition and entanglement. The fundamental principles
of building up quantum computation as well as experimental realizations of a quantum computer have been
carefully studied in the past decades.  The textbook  \cite{NC2011} has a pedagogical and comprehensive introduction
to such interdisciplinary research field.

Solving the dynamic equation of a large-scale quantum system is a major obstacle
in numerical simulation to overcome. So how to perform  a large-scale  quantum computing remains a
big challenge for a long term. Nevertheless, an integrable system is exactly solvable. If a quantum
computer is an integrable system with control terms, then it seems reasonable to devise appropriate methods
to perform information processing tasks. Hence  a realistic quantum computer
has been naturally assumed to be associated with quantum integrable systems \cite{Zhang11}.

In fault-tolerant quantum computation \cite{NC2011}, quantum information processing is robust against unavoidable noise.
Consider such noise as small perturbation. The classical KAM theorem \cite{Arnold99, GPS01} shows that the action-angle
variables of a classical non-degenerate integrable system are only slightly changed under sufficiently small perturbation.
The quantized version of the classical KAM theorem is called the quantum KAM theorem \cite{RL87} which tells that an
associated quantized integrable system is still stable under small perturbation. Thus quantum computation using action
variables is regarded as fault-tolerant quantum computation, whose fault-tolerance is justified by the quantum KAM theorem.

Recently, Lloyd and Montangero \cite{LM14} have proposed that universal quantum computing can be performed in integrable systems
with a global control field. They represent a qubit by  quantum action variables of integrable systems and construct
universal quantum gates via resonant quantum control. In this paper, we carry out a further research on quantum computation using
action variables. Remarkably, we introduce the crucial ideas of integrable systems into the field of quantum computation and quantum
information, including the quantum Hamilton--Jacobi equation \cite{LP83}, the quantum KAM theorem \cite{RL87} and the Birkhoff normal
forms \cite{Arnold99}. About mathematics of action-angle variables, we refer to Arnold's textbook \cite{Arnold99}; about physics
and notation, we refer to Goldstein's \cite{GPS01}.

The paper is organized as follows. Section 2 reviews basic concepts of the Liouville integrable system, including classical action
variables in central potential fields as well as the classical KAM theorem. Section 3 uses the quantum Hamilton--Jacobi equation to
calculate quantum action variables in central potential fields. Remarks on the application of the quantum KAM theorem to
fault-tolerant quantum computation are made. Section 4 presents a detailed perturbative study on the resonant driving theory in
the Dyson series. Section 5 shows the construction of universal quantum gates in central potential fields. Section 6 discusses an
extended harmonic oscillator quantum computation in the Birkhoff normal forms. Section 7 suggests a unified description for quantum
computation in integrable systems. Section 8 concludes with comments on further research.


\section{Action variables in classical integrable systems}

In this section, we make a brief introduction to the action-angle variables of a classical integrable system \cite{Arnold99, GPS01, Born27, Desloge82},
consisting of the Liouville definition, the Hamilton-Jacobi equation and the classical KAM theorem. Then we study integrable models in central
force fields, including the harmonic oscillator, the anharmonic oscillator \cite{Born27}, the Coulomb potential, and the Coulomb potential with
a perturbation term \cite{Desloge82}.

\subsection{The Liouville integrable system}

A Hamiltonian system of $k$ degrees of freedom has independent generalized coordinates $q_i$ and momenta $p_i$ with $i=1, \cdots, k$.
The coordinates of its $2k$-dimensional phase space are simply denoted by canonical variables $(q, p)$, where $q$ denotes the set of $q_i$ and
$p$ does of $p_i$. A physical quantity is a function of $(q,p)$ and time $t$; for example, the Hamiltonian of the system is $H_0=H_0(q,p,t)$.
When the Poisson bracket between two physical quantities vanishes, they are called commutative. When a  physical quantity is commutative with the
Hamiltonian, it is called conserved.

A classical system of $k$ degrees of freedom is called a Liouville integrable system if it has $k$ independent commutative conserved
quantities. In accordance with \cite{Arnold99}, the canonical variables $(q,p)$ of the Liouville integrable system
can be reformulated as the action-angle variables $(w,J)$, where the angle variables $w$ denote the set of $w_i$ and the action variables $J$ does of
$J_i$. The Hamiltonian $H_0$ is a function of action variables, $H_0=H_0(J)$, and so the action variables $J_i$ satisfy
\eq
  \{H_0,J_i\}=0, \quad  \{J_i,J_j\}=0,\qquad i,j=1,2,\cdots, k.
\en
  In the action-angle variable approach,
the conserved Hamiltonian $H_0$ directly gives rise to the frequency of the periodic motion by $\nu^c_i=\frac{\partial H_0}{\partial J_i}$,
without solving the equation of motion in the routine approach, so that the angle variable $w_i$ has the form $w_i(t)=\nu^c_i t+ w_i(0)$.

The constant action variables $J_i$ restrict the motion of the system into a $k$-dimensional subspace of the $2 k$-dimensional phase space. This subspace is
called a $k$-dimensional invariant torus $T^k$ labelled by angle variables, $T^k=\{(w_1,\cdots,w_k)\, \textrm{mod} \, 2 \pi\}$. Such a motion on $T^k$ is
called a conditionally periodic motion.

\subsection{Non-degenerate integrable system}

An integrable system is called non-degenerate if there exists a non-vanishing determinant,
\eq\label{non-degenerate_condition}
\det{\left(\frac{\partial^2 H_0}{\partial J_i\partial J_j}\right)}\neq 0;
\en
otherwise, it is degenerate. In a non-degenerate integrable system, classical frequencies are rationally independent in the almost entire phase
space (except regions of measure zero). Note that the classical frequencies
$\nu^c_1$, $\nu^c_2$, $\cdots$,  $\nu^c_k$ are called rationally independent (or incommensurate or non-resonant) if a linear combination of such frequencies
with integer coefficients $l_1$, $l_2$, $\cdots$, $l_k$, given by the equation
\eq
l_1 \nu^c_1 +l_2 \nu^c_2 +\cdots +l_k \nu^c_k=0,
\en
has no solution except all $l_i$ of zeros. The trajectory of an integrable system with rationally independent frequencies is dense everywhere and fills the whole
phase space (the torus $T^k$), so that a unique phase space can be specified.

\subsection{The Hamilton-Jacobi equation}

The action-angle variables $(w,J)$ can be calculated with the Hamilton-Jacobi equation.
A completely separable integrable system paints a periodic trajectory in the phase space. Its time-independent Hamiltonian $H_0(q,p)$
is the total energy $E_0$ of the system. The canonical transformation from $(q,p)$ to $(w,J)$ is generated by the Hamilton characteristic function
$W(q,J)=\sum_i W_i(q_i,J)$. The transformation equations are
\eq
 p_i=\frac {\partial W_i(q_i,J)} {\partial q_i}, \quad w_i=\frac {\partial W(q,J)} {\partial J_i},
\en
so that the Hamilton-Jacobi equation has the form
\eq
H_0 \left(q_1,\cdots,q_k;\frac{\partial W}{\partial q_1},\cdots,\frac{\partial W}{\partial q_k}\right)=E_0,
\en
with constants of integration $\alpha_2, \cdots, \alpha_k$.

For convenience, specify $H_0(q,p)=\alpha_1$ and denote the set of constants $\alpha_1,\alpha_2, \cdots, \alpha_k$ by $\alpha$.  Suppose the constant action
variable $J_i$ as an invertible function of constants $\alpha$, namely $J_i=J_i(\alpha)$, and thus $\alpha_i=\alpha_i(J)$. Hence the generating function
is $W(q,\alpha)\equiv W(q,J(\alpha))$ and the generalized canonical momenta are $p_i=p_i(q_i,\alpha)$. Interpret $J_i$ as an invariant area that the system depicts in the
phase space of $(q_i,p_i)$ at a complete period. That is, $J_i$ is a line integral over a period in the phase space,
$J_i=\oint p_i\,d q_i$, where $J_i=J_i(\alpha)$ is automatically satisfied.

\subsection{Classical integrable systems in central force fields} \label{classical central potential}

Choose the spherical polar coordinates $(r,\theta,\varphi)$ with radius $r$, polar angle $\theta$ and azimuthal angle $\varphi$.
Focus on the motion of a particle with mass $m$ in the central force field $V(r)$.  The $V(r)$ only depends on the radial distance $r$
from the centre of the field to particle. Denote action variables $(J_r, J_\theta, J_\varphi)$ and classical frequencies
$\omega^c_i=2\pi \nu^c_i$ where $i=r, \theta, \varphi$.

Consider a three-dimensional isotropic harmonic oscillator with $V(r)=\frac{1}{2}m\omega^2r^2$ where  $\omega$ is constant
frequency.  The Hamiltonian $H_0$, or energy $E_0=H_0$, is expressed as
\eq
\label{HSO energy}
H_0 =\frac{\omega}{2\pi}\left(2 J_r+J_\theta+J_\varphi\right).
\en
The classical frequencies $\omega^c_r$, $\omega^c_\theta$ and $\omega^c_\varphi$ respectively given by
\eq
\omega^c_r=2 \omega ;\quad \omega^c_\theta=\omega^c_\varphi=\omega,
\en
are obviously commensurate, and hence the isotropic harmonic oscillators are degenerate integrable models.

For  a three-dimensional anharmonic oscillator with a perturbation term \cite{Born27},
\eq
V(r)=\frac{1}{2}m\omega^2(r^2+cr^4),
\en
where $c$ is a sufficiently small positive constant, the Hamiltonian $H_0$ has the form
\eq
\label{AHO energy}
H_0=\frac{2m\omega^2}{3c}\left(1-\sqrt{1-\frac{3}{2\pi m\omega}
   (2J_r+J_\theta+J_\varphi) c+\frac{3}{16 \pi^2 m^2 \omega^2}(J_\theta+J_\varphi)^2 c^2}\right),
\en
leading to the classical frequencies $\omega^c_r$ and $\omega^c_\theta=\omega^c_\varphi$  respectively given by
\eq
\label{AHO frequency}
\omega^c_r = 2 \omega\frac{1}{1-\frac{3\,E}{2m\omega^2}\, c}, \quad
\omega^c_\varphi =\omega \frac{1-\frac{1}{4\pi m\omega}(J_\theta+J_\varphi)\, c}{1-\frac{3\,E}{2m\omega^2}c}.
\en
It is obvious that both two-dimensional anharmonic oscillators in the $\{r,\theta\}$ or $\{r,\varphi\}$ plane and one-dimensional anharmonic
oscillators in the $r$ direction are non-degenerate.

Take the Coulomb potential, $V(r)=-\frac{k}{r}$, with positive constant $k$. If the energy $H_0$ is negative, the motion
of the system is multi-periodic. Then the energy $H_0$ is
\eq
H_0=-\frac{2\pi^2 mk^2}{(J_r+J_\theta+J_\varphi)^2}.
\en
All of the classical frequencies $\omega^c_r, \omega^c_\theta, \omega^c_\varphi$ are the same as
\eq
\omega^c_r=\frac{(-2 m E)^{3/2}}{m^2 k}.
\en
Therefore, the models in the three-dimensional and two-dimensional Coulomb potentials are degenerate, but the one-dimensional
model in the radial direction is non-degenerate.

Look at the Coulomb potential with a second-order correction term \cite{Desloge82},
\eq
V(r)=-(\frac{k}{r}+\frac{\beta}{r^2}),
\en
where a small positive parameter $\beta$ satisfies $\beta \ll k r$. The negative energy $H_0$ for a bound motion has the form
\eq
H_0=-\frac{2\pi^2 m k^2}{(J_r+\sqrt{(J_\theta+J_\varphi)^2-8\pi^2m\beta})^2},
\en
with the classical frequencies $\omega^c_r$ and $\omega^c_\theta=\omega^c_\varphi$ given by
\eq
\omega^c_r=\frac{(-2mE)^{3/2}}{m^2 k},\quad \omega^c_\varphi=\omega^c_\theta=\frac{J_\theta+J_\varphi}{(J^2_\varphi-8\pi^2m\beta)^{1/2}}\omega^c_r.
\en
Explicitly, two-dimensional models in the $\{r,\theta\}$ or $\{r,\varphi\}$ direction are non-degenerate, as well as
one-dimensional models in the $r$ direction.

\subsection{Classical KAM theorem}

Add a perturbation Hamiltonian $H_1$, together with a sufficiently small parameter $\epsilon\ll1$, to a non-degenerate integrable Hamiltonian $H_0$. The total Hamiltonian is
$H=H_0+\epsilon H_1$. The classical KAM theorem \cite{Arnold99} shows that almost all the allowed motions of the perturbed system $H$ are still confined on the invariant torus $T^k$. That is, the motion of the integrable system $H_0$ is relatively stable under small perturbations. The perturbed motion is still a conditionally periodic motion,
except that the frequencies are slightly changed.

Consider the application of the classical KAM theorem to a degenerate integrable system $H_0$. First of all, by combining a small
Hamiltonian $H_1$, the intermediate system becomes non-degenerate integrable. Then one introduces a perturbation Hamiltonian
$H_2$ into the intermediate system  so that the final perturbed system satisfies the conditions of the classical KAM theorem.¡¡ ¡¡

\section{Action variables in quantized integrable systems}

This section describes the main properties of quantized integrable systems for the sake of quantum computation. First, the relation of classical frequencies
to quantum frequencies is made clear under the Bohr-Sommmerfeld quantization condition \cite{Born27}. Second,  the quantum Hamilton--Jacobi equation \cite{LP83}
is used to calculate the quantized Hamiltonian in central force field. Third, the application of the quantum KAM theorem \cite{RL87} to fault-tolerant quantum
computation is briefly discussed.

\subsection{Rationally independent quantum frequencies}

Canonical quantization of a classical integrable system produces a quantized integrable system. It is characterized by a set of quantum action operators $\hat{J}_i$
and the Hamiltonian $\hat{H}_0$. They satisfy the commutators, $[\hat{H}_0,\hat{J}_i]=0$ and  $[\hat{J}_i,\hat{J}_j]=0$ with $i,j=1,2,\cdots, k$.
The Hamiltonian $\hat{H}_0$ is a functional of action operators: $\hat{H}_0=\hat{H}_0(\hat{J})$.  The eigenstates $|j\rangle \equiv |j_1,j_2,\cdots, j_k\rangle$
are solutions of the eigenvalue equations $\hat{J}_i|j\rangle=j_i|j\rangle$ with eigenvalues $j_i$; the energy eigenstates $|n\rangle\equiv |n_1,n_2,\cdots,n_k\rangle$
satisfy $\hat{H}_0|n\rangle=E_0(n)|n\rangle$ with energy eigenvalues  $E_0(n)$, where the number $n$ is just another label of $j$ (or $j_1, j_2, \cdots, j_k$).

Quantum frequencies are conceptually different from classical frequencies. Quantum frequency is specified as an energy difference between two energy levels of a
quantum system. Denote the energy eigenvalue $E_0(n)=\hbar \omega_n$ with frequency parameter $\omega_n$.  Take two close energy levels $E_0(n_i+1)$ and $E_0(n_i)$,
where other quantum action numbers $n_j$ at $j\neq i$ are the same. So the quantum frequency is $\omega_{n_{i+1}n_i}=\omega_{n_{i+1}} -\omega_{n_i}$.  Here
denote $\omega_i \equiv \omega_{n_{i+1}n_i}$ for simplicity.

The Bohr-Sommerfeld quantization condition \cite{Born27} shows $j_i =2 \pi \hbar n_i$ with the Planck constant $\hbar$ and integer numbers $n_i$.
Applying  it to a classical Hamiltonian $H_0$ induces a quantized Hamiltonian $\widetilde{H}_0$. The quantum analogue of classical frequency is thus given
by $\widetilde{\omega}^c_i={\frac 1 \hbar}\frac{\partial {\widetilde {H}}_0}{\partial n_i}$.  The relation between two frequencies $\omega_i$
and $\widetilde{\omega}^c_i$ is given by
\eq
\omega_i =\widetilde{\omega}_i^c +\sum_{l=2}^\infty \frac 1 {l!} \frac {\partial^l {\widetilde H}_0 } {\hbar^l \partial n_i^l}.
\en
Obviously, the frequency $\omega_i$ may approximate the classical frequency $\tilde{\omega}^c_i$ very well at large action variables $j_i$ (or $n_i$). For example \cite{Born27},
one has $\omega_i =\widetilde{\omega}_i^c$ in three dimensional isotropic harmonic oscillators, and $\omega_i=\widetilde{\omega}_i^c$ in the Coulomb potential at large $n_i$.

Since rationally independent quantum frequencies are required in the selective driving theory \cite{LM14}, it is significant to investigate whether quantum frequencies in integrable systems are rationally independent or not. For example, classical frequencies in a non-degenerate classical integrable system are non-resonant, so that quantum frequencies in the quantized integrable system can be incommensurate at large action variables.

\subsection{The quantum Hamilton--Jacobi equation in central force field}

The quantum Hamilton--Jacobi equation \cite{LP83}, a fundamental topic in quantum mechanics, defines a quantization procedure obviously
different from ones in textbooks. It can be explored in the Bohm hidden-variable approach to quantum mechanics, and in particular
can be investigated in high energy physics, such as quantum chromodynamics. Calculation in various physical models had been performed to verify
that the quantum Hamilton--Jacobi equation not only yields previous results but also creates something novel.

Suppose a quantum particle of mass $m$ in a central potential $V(r)$. The Hamiltonian $\hat{H}_0$ has the form
\eq
\hat{H}_0=-\frac{\hbar^2}{2m}\left[\frac{1}{r^2}\frac{\partial}{\partial r}\left(r^2\frac{\partial}{\partial r}\right)+\frac{1}{r^2\sin\theta}\frac{\partial}{\partial\theta}(\sin\theta\frac{\partial}{\partial\theta})
+\frac{1}{r^2}\frac{1}{\sin^2\theta}\frac{\partial^2}{\partial\varphi^2}\right]+V(r).
\en
The time-independent Schroedinger equation is $\hat{H}_0 \psi_{E_0}(\vec{r})=E_0 \psi_{E_0}(\vec{r})$ with
energy eigenvalue $E_0$ and wavefunction $\psi_{E_0}(\vec{r})$.

The Bohr--Sommerfeld quantization condition tells us that the action variables $j_r$, $j_\theta$ and $j_\varphi$
are quantized as
\eq
 j_r = 2 \pi \hbar n_r, \quad j_\theta = 2 \pi \hbar n_\theta, \quad  j_\varphi = 2 \pi \hbar n_\varphi,
\en
with integer numbers $n_r$, $n_\theta$ and $n_\varphi$. Thus the Hamiltonian $H_0$ (or the energy $E_0$) is a function of quantized action
variables  $j_r$,  $j_\theta$, $j_\varphi$ (or $n_r$, $n_\theta$ and $n_\varphi$).

With the quantum Hamilton characteristic function
\eq
W({\vec r},E_0)=W_r(r)+W_\theta(\theta)+W_\varphi(\varphi),
\en
the wavefunction is reformulated as $\psi_{E_0}(\vec{r})=e^{\frac{i}{\hbar} W(\vec{r},E_0)}$. Define the quantum
momenta $p^\ast_r$, $p_\theta^\ast$  and $p^\ast_{\varphi}$ respectively as
\eq
 p_r^*=\frac{\partial W_r}{\partial r}, \quad p_\theta^*=\frac{\partial W_\theta}{\partial \theta},
\quad p_\varphi^*=\frac{\partial W_\varphi}{\partial \varphi}.\en
Calculate the eigenvalues of quantum action variables in the spherical coordinates,
\eq
j_r =\oint_{D_r} p^*(r)d r, \quad j_\theta =\oint_{D_\theta} p^*(\theta)d\theta,\quad j_\varphi =\oint_{D_\varphi} p^*(\varphi) d\varphi,
\en
where $D_r$, $D_\theta$ and $D_\varphi$ are closed contours in the complex plane and they are respectively chosen by the poles of integrands in classical physics \cite{LP83}.

By the method of separation of variables, the time-independent Schroedinger equation in terms of $p^\ast_r$, $p_\theta^\ast$ and  $p^\ast_{\varphi}$ are separated into
three independent quantum Hamilton-Jacobi equations, respectively called the azimuthal angular equation, the polar angular equation and the radial equation. Solve these
equations for quantum momenta and then calculate quantum action variables.

Solve the azimuthal angular equation in terms of the quantum momentum $p^*(\varphi)$,
\eq
-i \hbar\frac{\partial p^*(\varphi)}{\partial\varphi}+p^{*2}(\varphi)=\alpha^2_\varphi, \label{varphi_ricatti_equation}
\en
with constant $\alpha_\varphi$, so we have the action variable $j_\varphi=2\pi\alpha_\varphi$, namely $\alpha_\varphi=n_\varphi\hbar$. Solve the polar angular
equation with constant $\alpha_\theta$,
\eq
i \hbar\frac{\partial p^*(\theta)}{\partial\theta}+p^{*2}(\theta)
-i \hbar\frac{\cos\theta}{\sin\theta}p^*(\theta)+\frac{\alpha^2_\varphi}{\sin^2\theta}
=\alpha^2_\theta,\label{theta_ricatti_equation}
\en
for the quantum momentum $p^\ast(\theta)$, and calculate $j_\theta$ to obtain $\alpha^2_\theta=\hbar^2 l(l+1)$, with $l=n_\theta+n_\varphi$, $l=0,1,2,\cdots$,
which is the angular momentum in the Schroedinger equation approach to quantum mechanics. The radial equation in terms of $p^\ast(r)$ is given by
\eq
  -i \hbar \frac{\partial p^*(r)}{\partial r}
+p^{*2}(r) - 2i \hbar \frac{p^*(r)}{r}  +\frac{\alpha^2_\theta}{r^2}= 2 m (E-V(r)),  \label{r_ricatti_equation}
\en
which depends on the symmetric spherical potential $V(r)$.

Let us concentrate on $V(r)$ listed in Subsection~\ref{classical central potential}.
When the central potential $V(r)$ describes a three-dimensional anharmonic oscillator with a small perturbation
term $cr^4$,  solve the radial equation and calculate the contour integral of defining $j_r$ to obtain the quantized Hamiltonian,
\eq
H_0=\frac {2 m \omega^2} {3 c} \left( 1-\sqrt{ 1-  \frac {3 \hbar } { m \omega} (2 n_r + l+ \frac 3 2)c
 + \frac {3 \hbar^2} {4 m^2 \omega^2 }(l+\frac 3 2 )(l- \frac 1 2 ) c^2 } \right)
\en
which has an approximation form at small $c$,
\eq
 H_0=\hbar\omega(2n_r+l+\frac{3}{2})-\frac{\hbar^2}{4m} (l+\frac{3}{2})(l-\frac 1 2)\, c.
 \en
Obviously£¬ at $c=0$, we have the quantized Hamiltonian for a three dimensional isotropic harmonic oscillator:
$H_0=\hbar\omega(2n_r+n_\theta+n_\varphi+3/2)$. When the $V(r)$ is the Coulomb potential with second-order perturbation term, the quantized
Hamiltonian has the form
\eq
H_0=-\frac{mk^2}{2(n_r+l'+1)^2\hbar^2},
\en
with $l'=-\frac{1}{2}+\sqrt{(l+\frac{1}{2})^2-\frac{2m\beta}{\hbar^2}}$. At $\beta=0$, we have the well known quantized Hamiltonian in
the Coulomb potential,
\eq
H_0=-\frac{mk^2}{2(n_r+n_\theta+n_\varphi+1)^2\hbar^2}.
\en

The quantum frequencies between two close energy levels in a central potential field are respectively denoted by $\omega_r$, $\omega_\theta$ and $\omega_\varphi$.
In a three-dimensional harmonic oscillator, we have $\omega_r =\omega$ and $\omega_\theta=\omega_\varphi=\omega$. In the three-dimensional anisotropic harmonic
oscillator, we have $\omega_r \approx 2 \omega$ at small $c$ but $\omega_\theta \neq \widetilde{\omega}^c_\theta$ at large action variables. In the three-dimensional
Coulomb potential, we have $\omega_r=\omega_\theta=\omega_\varphi$ with
\eq
\omega_r=\frac {m k^2} {2 \hbar^2} \frac {2 n+ 3} {(n+1)^2 (n+2)^2}, \quad \widetilde{\omega}^c_r= \frac {m k^2} {\hbar^2 (n+1)^3},
\en
where $n=n_r+n_\theta+n_\varphi$, so that $\omega_r \approx \widetilde{\omega}^c_r$ at large $n$. In the above Coulomb potential with small perturbation, we have
$\omega_r \neq \omega_\theta$ but $\omega_\theta =\omega_\varphi$.

Note that a quantum degenerate system in quantum mechanics has at least two linearly independent eigenvectors  for a given energy eigenvalue.
Here is a natural question of whether canonical quantization of a classical degenerate integrable system gives rise to a quantum degenerate system.
The answer is obvious for the above quantized
integrable systems in central potential fields. In contrast, it is worthwhile thinking about whether canonical quantization of a classical non-degenerate integrable
system induces a quantum non-degenerate integrable system. The reason of asking such questions is that quantum computation in a quantum degenerate system
can be quite different from one in a quantum non-degenerate system.

\subsection{Quantum KAM theorem and fault-tolerant quantum computation}
\label{QKAM_FTQC}

A quantum composite system consists of a free system and a control system. The free Hamiltonian $\hat{H}_0$ possesses an energy spectrum of $E_0(n)$ with eigenstates $|n\rangle$.
The control Hamiltonian $\hat{H}_1$, in the orthonormal basis $|n\rangle$, is formally expressed as
\eq
\label{control hamiltonian}
\hat{H}_1 =\hbar \sum_{n,n'} a_{n n'} |n\rangle\langle n'|,
\en
with amplitudes $a_{nn'}$. So the total Hamiltonian is $\hat{H}=\hat{H}_0 + \epsilon(t) \hat{H}_1$ with a time-dependent control parameter $\epsilon(t)$.
As in the resonant quantum control for quantum computation \cite{LM14}, $\epsilon(t)$ is given by
$\epsilon(t)=\epsilon\, \cos(\omega_{d} t+\varphi)$ with real time-independent parameter $\epsilon$, driving frequency $\omega_{d}$ and initial phase $\varphi$.

When the free Hamiltonian $\hat{H}_0$ represents a quantum integrable system and the control Hamiltonian $\epsilon(t) \hat{H}_1$ is a small perturbation term, the dynamics of such
a composite system can be characterized by the quantum KAM theorem \cite{RL87}. Roughly speaking, the quantum KAM theorem, as a quantum generalization of the classical
KAM theorem, exhibits the stability of a quantum integrable system under small perturbation.  As various noise interacting with a quantum computer is regarded as small perturbation, one is allowed to investigate quantum computation using action variables from the viewpoint of fault-tolerant quantum computation \cite{NC2011}.

When the driving frequency $\omega_d$ is equal to frequency difference between two energy levels
of the free Hamiltonian, we have on-resonant quantum states.
Concerning quantum computation, on-resonant quantum states are used to encode logical qubits. Note that the contribution of off-resonant quantum states to a physical process
can be neglected. Hence quantum computation in integrable systems is argued to be fault-tolerant because
of the quantum KAM theorem. As the control Hamiltonian $\epsilon(t) \hat{H}_1$ is rather larger than the free Hamiltonian $\hat{H}_0$, however, quantum chaos has to be considered,
which is beyond the quantum KAM theorem.

Make the above argument clear. Consider the Coulomb potential. A chosen physical model is an electron moving around the hydrogen
atom nucleus or an electron of alkali metal at the outer shell. The free Hamiltonian is
\eq
\hat{H}_0 =\frac {\hat{{\vec p}}^2} {2 m} - \frac {Z e^2} {4 \pi \epsilon_0} \frac {1} {|\hat{\vec{r}}|},
\en
with spatial momentum operator $\hat{\vec{p}}$, electron mass $m$, electric charge $-e$, effective electric charge $Z e$ of the nucleon, dielectric constant $\epsilon_0$ and radial distance vector operator $\hat{\vec{r}}$. The control Hamiltonian  is given by the dipolar interaction $\epsilon(t)\hat{H}_1 =- \epsilon(t) \vec{E} \cdot \hat{\vec{r}}$ with classical electric field strength vector $\vec{E}$. As $\epsilon$ is rather small,  the quantum KAM theorem allows us to investigate fault-tolerant quantum computation. But as $\epsilon$ is larger enough, quantum chaotic behaviour \cite{LM14} may forbid quantum computation under global control.

For a physical model of charged particles in the Coulomb potential interacting with a classical oscillating electric field,  one builds up universal quantum computation via
the resonance method \cite{NC2011}. First of all, the resonant quantum control in the Coulomb potential can be physically simplified as the JC model in quantum optics, where the classical electric field is quantized as a collection of photons in quantum field theory. It is the JC model that takes the responsibility for  quantum computation via cavity quantum electrodynamics and via ion trap approach. In this respect, quantum computation using action variables in the paper is as meaningful and powerful as known experimental quantum computation schemes.  Remarkably, the fault-tolerance of quantum computation in integrable systems, namely the fault-tolerance of the JC model in quantum computation, can be investigated in the scope of the quantum KAM theorem.

Note that what we have done in this paper has no obvious connection with quantum computation using spin degrees of freedom, such as the ion trap quantum computation and the
nuclear magnetic resonance (NMR) quantum computation. In quantum computation using spins, a spin-1/2 Hilbert space acts as a physical qubit, while quantum control via magnetic dipole interaction naturally yields single-qubit gates and two-qubit gates. After all, the fault-tolerance of quantum computation in spin integrable systems has to be investigated in accordance with the quantum KAM theorem in spin systems. Hence the study of quantum computation using action variables in spin systems is helpful to understand the NMR quantum computation and the ion trap quantum computation.

So far the description of the quantum KAM theorem \cite{RL87}  has been not completely fixed yet partly because a quantization procedure from a classical theory to a quantum theory is not unique. Therefore, the study of fault-tolerant quantum computation in integrable systems using action variables sheds a light on an appropriate characterization of the quantum KAM theorem. For example, the quantum KAM-like theorem states that  the energy eigenstate $|n'\rangle$ of the total Hamiltonian can be well described by quantum action variable numbers in the free integrable system, provided that the quantum localization condition \cite{HT83}: $|\langle  n'|n \rangle|^2 >1/2$, is satisfied. Therefore it is possible to make clear the interpretation of the quantum localization condition from the viewpoint
of fault-tolerant quantum computation.

\section{A study of resonant quantum control for quantum computation}

Universal quantum computation in quantum circuit model \cite{NC2011} can be performed by a universal quantum gate set consisting of single-qubit gates and a nontrivial
two-qubit gate. We investigate the control of quantum systems for the construction of quantum gates in detail, so that the preliminary and necessary knowledge for
the future study of quantum computation using action variables is well prepared. First of all, we use the Dyson series to derive the effective unitary evolution
operator for an on-resonant transition. Then we construct single-qubit gates in a two-level Rabi oscillation model as well as two-qubit gates in a four-level system by
the method of frequency selection. Afterwards, we close with remarks on further research.

\subsection{The Dyson series of the unitary evolution operator}

Consider a composite system of a free system and a control system as in Subsection \ref{QKAM_FTQC}. Take energy eigenvectors $|n\rangle$ and $|n'\rangle$ with energy
eigenvalues $E_0(n)>E_0(n')$, and define the frequency parameters $\omega_n$ and $\omega_{n'}$ by $E_0(n)=\hbar \omega_n$ and $E_0(n')=\hbar \omega_{n'}$, so that
the positive frequency  is $\omega_{nn'}=\omega_n-\omega_{n'}$.

Start with a quantum state $|\psi\rangle$ in the Schrodinger picture.  The free unitary time evolution operator is $\hat{U}_0(t)=e^{-i \hat{H}_0 t/\hbar}$.
Define a quantum state $|\psi_I\rangle$ in the interaction picture as $|\psi_I\rangle=\hat{U}_0^\dag|\psi\rangle$. The time evolution of $|\psi_I\rangle$ is
determined by the interaction Hamiltonian $\hat{H}^I=\epsilon(t)\hat{U}_0^\dag \hat{H}_1 \hat{U}_0$.
The $\hat{H}^I$ is a sum of three terms:
\eq
\hat{H}^I(t)=\hat{H}^I_{+}(t) + \hat{H}^I_{-}(t) + \hat{H}^I_{g}(t).
\en
The $\hat{H}^I_+(t)$ and $\hat{H}^I_{-}$ are respectively given by
\eq
\label{low hamiltonian}
\hat{H}^I_{\pm}  = \frac{\hbar\epsilon }{2} \sum_{E_0(n)> E_0(n')}\left(  {a_{nn'} e^{\pm i\varphi}e^{it(\omega_{nn'} \pm \omega_d)}|n\rangle\langle n'| +c.c.}\right),
\en
and the $\hat{H}^I_g(t)$ is given by
$\hat{H}^I_{g} = \hbar\epsilon \cos(\omega_d t+\varphi) \sum_{n} a_{nn}|n\rangle\langle n|$.

The time evolutional operator $\hat{U}^I(t)$ has the form of the Dyson series,
\eq \hat{U}^I(t)=\sum_{n=0}^{\infty} \hat{U}^{I(n)}(t), \en
where the $n$th order term $\hat{U}^{I(n)}(t)$ is recursively defined as
\eq
\label{dyson_recursive}
\hat{U}^{I(n)}(t)=-\frac {i} {\hbar} \int^t_0 \hat{H}^I(t_1) \hat{U}^{I(n-1)}(t_1) d t_1,
\en
with the zero'th order term $\hat{U}^{I(0)}=1$.

\subsection{A perturbative study of on-resonant transition}

From an $N$-level free quantum system, $n=1,2,\cdots,N$, we wish to select an on-resonant two-level
Rabi oscillation subsystem, spanned by energy eigenstates $|m\rangle$ and $|m'\rangle$ with energy eigenvalues $E_0(m) > E_0(m')$. Set the driving frequency
$\omega_d=\omega_{mm'}$. We will verify $\hat{U}^I(t)=\hat{U}^I_{Rabi}(t)$ under suitable constraints with the resonant driving theory. The unitary evolutional
operator $\hat{U}^I_{Rabi}(t)$, determined by the time-independent Rabi oscillation Hamiltonian $\hat{H}^I_{Rabi}=\hat{H}^I_{-}|_{\omega_d=\omega_{mm'}}$,
has the form
\eq\label{rabi evolution_operator}
\hat{U}^I_{\small Rabi}(t)=\exp\left(-\frac{i\epsilon t}{2}(a_{mm'}e^{-i\varphi}|m\rangle\langle m'|+c.c.)\right),
\en
which is used in the construction of quantum gates.

To first order, the $\hat{U}^I(t)$ is the summation of $\hat{U}_+^{I(1)}(t)$,  $\hat{U}_-^{I(1)}(t)$ and $\hat{U}_g^{I(1)}(t)$ given by
\eq
\hat{U}_\pm^{I(1)}(t)=(-i/\hbar) \int^t_0 \hat{H}_\pm^I(t_1) d t_1, \qquad \hat{U}_g^{I(1)}(t)=(-i/\hbar) \int^t_0 \hat{H}_g^I(t_1) d t_1.
\en
After calculation,
the $\hat{U}_+^{I(1)}(t)$ is approximately a polynomial of $\epsilon |a_{nn'}|/(\omega_{nn'}+\omega_d)$, namely
$\hat{U}^{I(1)}_{+}(t)\approx \sum_{E_0(n) > E_0(n')}\mathcal{O}\left(\frac{\epsilon |a_{nn'}|}{\omega_{nn'}+\omega_d}\right)$.
Similarly, $\hat{U}^{I(1)}_{g}(t) \approx \sum_{n}\mathcal{O}\left(\frac{\epsilon |a_{nn'}|}{\omega_d}\right)$.
At $\omega_d=\omega_{mm'}$, the $\hat{U}_-^{I(1)}(t)$ is approximately expressed as
\eq
\hat{U}_-^{I(1)}(t)\approx\sum_{n n'\neq m m'}\mathcal{O}\left(\frac{\epsilon |a_{nn'}|}{\omega_{nn'}-\omega_{d}}\right)+ \hat{U}_{Rabi}^{I(1)}(t),
\en
where  $\hat{U}^{I(1)}_{Rabi}(t)=-i\hat{H}^I_{Rabi} t/\hbar$ leads to $\hat{U}^{I(1)}_{Rabi}(t) \approx \mathcal{O} (\epsilon |a_{mm'}|) t $.

Assume that $\omega_d$, $\omega_{nn'}$ and $|\omega_{nn'}-\omega_{d}|$ at $nn'\neq mm'$ are rather larger quantities compared to $\epsilon |a_{nn'}|$. That is,
$\epsilon |a_{nn'}|/(\omega_{nn'}+\omega_d)$, $\epsilon |a_{nn'}|/\omega_d$ and $\epsilon |a_{nn'}|/(\omega_{nn'}-\omega_{d})$ are sufficiently small. Meanwhile,
suppose that the evolutional time $t$ is also larger compared to $1 /(\omega_{nn'}+\omega_d)$, $1 /\omega_d$ and $1/(\omega_{nn'}-\omega_{d})$. Under such
constraints, we have $\hat{U}^{I(1)}(t)=\hat{U}^{I(1)}_{Rabi}(t)$.

We make a sketch of the proof for $\hat{U}^I(t)=\hat{U}^I_{Rabi}(t)$. Assume that $\hat{U}^{I(n)}(t)=\hat{U}^{I(n)}_{Rabi}(t)$ has been verified under
relevant constraints. The $\hat{U}^{I(n)}_{Rabi}(t)$, the $n$th-order term of the Rabi oscillation evolutional operator, has the form
\eq
\hat{U}^{I(n)}_{Rabi}(t)=(-i/\hbar)^n \frac {t^n} {n!} (\hat{H}^I_{\small Rabi}(t))^n.
\en
Replace $\hat{U}^{I(n)}(t)$ with $\hat{U}^{I(n)}_{Rabi}(t)$ in $\hat{U}^{I(n+1)}(t)$. Applying the integral formula
\eq
\int^t_0 \frac{(t_1)^n} {n!} e^{i(\omega_{nn'}\pm \omega_d) t_1} d t_1 \approx \mathcal{O}(\frac 1 {\omega_{nn'}\pm \omega_d})
\en
and the other formula $\int^t_0 \frac{(t_1)^n} {n!} \cos(\omega_d t_1+\varphi) d t_1 \approx \mathcal{O}(1/\omega_d)$, we verify
 $\hat{U}^{I(n+1)}=\hat{U}^{I(n+1)}_{Rabi}$ under on-resonant conditions.

The above calculation in the Dyson series explains the frequency selection of a two-level Rabi oscillation system from an $N$-level quantum system.
At $N=2$, we have actually presented a rigorous proof on the rotating-wave approximation \cite{LM14}. At $N>2$, however, we have to perform a more careful study
so as to determine which necessary constraints on frequencies and time are imposed.

\subsection{The construction of single-qubit gates at $\omega_d \neq 0$}

Now we construct single-qubit gates \cite{NC2011} via the selective driving. First of all, the simplest single-qubit gates are the Pauli gates: $X=\sigma_x$,  $Y=\sigma_y$ and $Z=\sigma_z$.
A rotational gate $R_{\hat{n}}(\theta)$ around the $\hat{n}$-axis $\vec{e}_n$ about angle  $\theta$ is denoted by $R_{\hat{n}}(\theta) =e^{-i\frac{ \theta }{2} \sigma_n }$
where $\sigma_n$ is the projection of the Pauli-matrix vector $\vec{\sigma}=(\sigma_x,\sigma_y,\sigma_z)$ along the unit vector $\hat{n}=\vec{e}_n$.
All single-qubit gates can be expressed as a product of rotational gates around the $x$-axis and rotational gates around the $y$-axis. For example, the identity gate $I_2=R_x(0)$; the Pauli gates $X=i R_x(\pi)$, $Y=i R_y(\pi)$ and $Z=R_x(\pi)R_y(\pi)$; the Hadmard gate $H=i\, R_x(\pi)R_y(\frac{\pi}{2})$ and the phase gate
$S= e^{-i\frac{\pi}{4}}\, R_y(-\frac{\pi}{2})R_x(\frac{\pi}{2})R_y(\frac{\pi}{2})$.

For simplicity, the energy eigenstates $|m'\rangle$ and $|m\rangle$ spanning the Hilbert space for the Rabi oscillation are respectively denoted as the
computational basis states $|0\rangle_L$ and $|1\rangle_L$. Thus the free Hamiltonian is $\hat{H}_0=\tilde{E}_0(0) |0\rangle_L \langle 0| + \tilde{E}_0(1) |1\rangle_L\langle 1|$ with $\tilde{E}_0(1)>\tilde{E}_0(0)$. It is reformulated as $\hat{H}_0=\frac 1 2 \hbar \overline{\omega}_{10} I_2 - \frac 1 2 \hbar \tilde{\omega}_{10} \sigma_z$ where
$\hbar \overline{\omega}_{10}=\tilde{E}_0(0) +\tilde{E}_0(1)$
and $\hbar \tilde{\omega}_{10}=\tilde{E}_0(1) - \tilde{E}_0(0)$. So the free evolution operator is
$\hat{U}_0(t)=e^{-{\frac i 2} \overline{\omega}_{10} t} R_{z}(-\tilde{\omega}_{10} t)$.
The interaction Hamiltonian $\hat{H}_1$ has the matrix form
\eq
\hat{H}_1=\hbar\left(
                   \begin{array}{cc}
                     \tilde{a}_{00} & \tilde{a}_{01} \\
                     \tilde{a}_{10} & \tilde{a}_{11} \\
                   \end{array}
                 \right),
\en
with $\tilde{a}_{10}=|\tilde{a}_{10}| e^{-i\tilde{\varphi}_{10}}$, and
the effective Hamiltonian $\hat{H}^I_{-}$ (\ref{low hamiltonian})  is given by
\eq
\hat{H}^I_{-} ={\frac 1 2}\hbar\epsilon |\tilde{a}_{10}| \left(\begin{array}{cc}
       0 & e^{i (\tilde{\Delta}_{10} t +\tilde{\varphi}_{10}')} \\
       e^{-i(\tilde{\Delta}_{10} t+\tilde{\varphi}_{10}') } & 0 \\
       \end{array}\right),
\en
where $\tilde{\varphi}_{10}'=\varphi+\tilde{\varphi}_{10}$ and $\tilde{\Delta}_{10}=\omega_d-\tilde{\omega}_{10}$.

At the on-resonant transition condition $\tilde{\Delta}_{10}=0$, the unitary evolutional operator for the Rabi oscillation in
the interaction picture is $\hat{U}^I_{Rabi}(t)=R_{n'_{10}}(\epsilon |\tilde{a}_{10}| t)$ with the unit vector
$\hat{n}_{10}'=(\cos\tilde{\varphi}'_{10},-\sin\tilde{\varphi}_{10}',0)$. At $\tilde{\varphi}_{10}'=0$, $\hat{U}^I_{Rabi}(t)=R_{x}(\epsilon |\tilde{a}_{10}| t)$
and at $\tilde{\varphi}_{10}'=-\frac \pi 2$, $\hat{U}^I_{Rabi}(t)=R_{y}(\epsilon |\tilde{a}_{10}| t)$.

We turn to the Schroedinger picture at $\tilde{\Delta}_{10}\neq 0$ to verify that the Rabi oscillation indeed occurs at $\tilde{\Delta}_{10}=0$.
The Hamiltonian in the Schroedinger picture is $\hat{H}_{-}=\hat{U}_0 \hat{H}^I_{-} \hat{U}_0^\dag$. Then go to the rotating frame picture
by $|\psi_R(t)\rangle=R(t)|\psi(t)\rangle$ with the rotation operator $R(t)=R_z(\omega_d t +\varphi)$. In the rotating frame picture,
the time independent Hamiltonian  $\hat{H}^R$  has the form
\eq
 \hat{H}^R=  \frac 1 2 \hbar (\overline{\omega}_{10} I_2 + \tilde{\Omega}_{10} \sigma_{n_{10}}),
\en
where  $\sigma_{n_{10}}$ and $\tilde{\Omega}_{10}$ are respectively given by
$\sigma_{n_{10}}=\vec{\sigma}\cdot \hat{n}_{10}$ with $\hat{n}_{10}=\frac {\vec{n}_{10}} {|\vec{n}_{10}|}$ and
the vector $\vec{n}_{10}=(\frac 1 2 \epsilon |\tilde{a}_{10}|,0,\frac 1 2 \tilde{\Delta}_{10})$ with $|\vec{n}_{10}|=\frac 1 2 \tilde{\Omega}_{10}$ and $\tilde{\Omega}_{10}=\sqrt{\tilde{\Delta}_{10}^2+\epsilon^2|\tilde{a}_{10}|^2}$,
so that the evolution operator is
\eq
 \hat{U}_R(t)=e^{-{\frac i 2} \overline{\omega}_{10} t} R_{n_{10}}(\tilde{\Omega}_{10} t).
\en
Therefore, the evolution operator in the Schroedinger picture,  $\hat{U}(t)=R^\dag(t) \hat{U}_R(t) R(0)$, has the matrix form
\eq
\hat{U}(t)=e^{-{\frac i 2} \overline{\omega}_{10} t} \left(
      \begin{array}{cc}
       e^{{\frac i 2} \tilde{\omega}_{10} t}(\cos\frac{\tilde{\Omega}_{10} t}{2}-i\frac{\tilde{\Delta}_{10}}{\tilde{\Omega}_{10}}\sin\frac{\tilde{\Omega}_{10} t}{2})
       & -i e^{{\frac i 2}  \tilde{\omega}_{10} t} e^{i\tilde{\varphi}_{10}'} \frac{\epsilon |\tilde{a}_{10}|}{\tilde{\Omega}_{10}}\sin\frac{\tilde{\Omega}_{10} t}{2} \\
      -i e^{-{\frac i 2}  \tilde{\omega}_{10} t} e^{-i\tilde{\varphi}_{10}'} \frac{\epsilon |\tilde{a}_{10}|}{\Omega_{10}}\sin\frac{\Omega_{10} t}{2}
        & e^{-{\frac i 2}  \tilde{\omega}_{10} t}(\cos\frac{\tilde{\Omega}_{10} t}{2}+i\frac{\tilde{\Delta}_{10}}{\tilde{\Omega}_{10}}\sin\frac{\tilde{\Omega}_{10} t}{2})\\
      \end{array}
    \right),\nonumber
\en
which leads to $\hat{U}_{Rabi}(t)=\hat{U}(t)|_{\tilde{\Delta}_{10}=0}$. Note that
$\hat{U}_{Rabi}(t)=\hat{U}_0(t) \hat{U}^I_{Rabi}(t)$.

Let us calculate the transition probability $P_{01}(\tilde{\Delta}_{10})=|{}_L\langle 1|\hat{U}(t)|0\rangle_L |^2$.  The result is
\eq
P_{01}(\tilde{\Delta}_{10})={\frac {|\epsilon|^2 |\tilde{a}_{10}|^2} {\tilde{\Omega}_{10}^2}} \sin^2{\frac{\tilde{\Omega}_{10} t} 2}.
\en
Obviously, when the driving frequency $\omega_d$ is the frequency difference $\tilde{\omega}_{10}$, namely $\tilde{\Delta}_{10}=0$, one has a two-level Rabi oscillation model
with $P_{01}=\sin^2{\frac{\tilde{\Omega}_{10} t} 2}$, so that the on-resonant transition probability $P_{01}=1$ occurs at $t={\frac \pi {\tilde{\Omega}_{10}}}$.

In addition, $\tilde{\Delta}_{10}=0$ with $\omega_d\neq 0$ tells $\tilde{E}_0(0)\neq \tilde{E}_0(1)$. That is, the Rabi oscillation
system is a two-level non-degenerate quantum system.

\subsection{The construction of single-qubit gates at $\omega_d =0$}

  At the driving frequency $\omega_d=0$, obviously, the above frequency selection technique cannot be exploited for the construction of single-qubit gates. Denote $\epsilon_0=\epsilon \cos\varphi$. The time-independent total Hamiltonian is $\hat{H}=\frac 1 2 \hbar(\omega_t I_2 + \tilde{\Omega}_0 \sigma_{n_0})$. The $\omega_t$
and $\omega_r$ are  \eq
\omega_t = \overline{\omega}_{10}+\epsilon_0 (\tilde{a}_{11}-\tilde{a}_{00}), \quad \omega_r = \tilde{\omega}_{10}+\epsilon_0 (\tilde{a}_{11}-\tilde{a}_{00}).
\en
The vector $\vec{n}_0=(\frac 1 2 \tilde{\gamma} \cos \tilde{\varphi}_{10}, -\frac 1 2 \tilde{\gamma} \sin \tilde{\varphi}_{10}, -\frac 1 2 \omega_r)$
with $\tilde{\gamma}=2 \epsilon_0 |\tilde{a}_{10}|$ defines the unit vector $\hat{n}_0=\frac {\vec{n}_0} {|\vec{n}_0|}$ and $|\vec{n}_0|=\frac 1 2 {\tilde{\Omega}_0}$ and $\tilde{\Omega}_0=\sqrt{\omega_r^2+\tilde{\gamma}^2}$ and $\sigma_{n_0}=\vec{\sigma}\cdot \hat{n}_0$.
The time evolutional operator in the Schroedinger picture, $\hat{U}_{\omega_d=0}(t)=e^{-{\frac i 2} \omega_{t} t} R_{n_0}(\tilde{\Omega}_0 t)$, has the form
\eq
\hat{U}_{\omega_d=0}(t)=e^{-{\frac i 2} \omega_t t} \left(
      \begin{array}{cc}
       \cos\frac{\tilde{\Omega}_0 t}{2}+i\frac{\omega_r}{\tilde{\Omega}_0}\sin\frac{\tilde{\Omega}_0 t}{2}
       & -i e^{i\tilde{\varphi}_{10}} \frac{2\epsilon_0 |\tilde{a}_{10}|}{\tilde{\Omega}_0}\sin\frac{\tilde{\Omega}_0 t}{2} \\
      -i e^{-i\tilde{\varphi}_{10}} \frac{2\epsilon_0 |\tilde{a}_{10}|}{\tilde{\Omega}_0}\sin\frac{\tilde{\Omega}_0 t}{2}
        &  \cos\frac{\tilde{\Omega}_0 t}{2} - i\frac{\omega_r}{\tilde{\Omega}_0}\sin\frac{\tilde{\Omega}_0 t}{2}\\
      \end{array}
    \right),
\en
from which the transition probability $P_{01}=|{}_L\langle 1|\hat{U}_{\omega_d=0}(t)|0\rangle_L |^2$ is clearly derived. At $\omega_r=0$, namely
$\tilde{\omega}_{10}=\epsilon_0 (\tilde{a}_{00}-\tilde{a}_{11})$, we have a two-level Rabi oscillation model with the unitary operator
 $\hat{U}_{\omega_r=0}(t)=e^{-{\frac i 2} \omega_{t} t} R_{n_0'}(2 \epsilon_0 |\tilde{a}_{10}| t)$ where the unit vector $\hat{n}_0'=(\cos\tilde{\varphi}_{10},-\sin\tilde{\varphi}_{10},0)$.

The  calculation for frequency selection in Subsection 4.2 requires the key quantity, $\epsilon_0 |\tilde{a}_{nn'}|/\omega_d$
sufficiently small at $\omega_d=\tilde{\omega}_{10}\neq 0$, whereas the quantity $\epsilon_0 |\tilde{a}_{nn'}|/{\tilde{\omega}_{10}} = |\tilde{a}_{nn'}|/(\tilde{a}_{00}-\tilde{a}_{11})$ at $\omega_d=0$  is not necessarily rather small.
Here is no explicit connection between the perturbative calculation at $\omega_d \neq 0$ and the non-perturbative case at $\omega_d=0$.
Therefore, it is interesting to study a non-perturbative construction of quantum gates at $\omega_d \neq 0$.

Additionally, at $\tilde{a}_{00} =\tilde{a}_{11}$, namely $\tilde{\omega}_{10}=0$,  a two-level degenerate quantum system is considered;
at $\tilde{a}_{00} \neq\tilde{a}_{11}$, a non-degenerate two-level quantum system is used.

\subsection{The construction of two-qubit gates at $\omega_d \neq 0$}

Typical two-qubit gates \cite{NC2011} in this paper include continuous SWAP gates  and control unitary gates.
The continuous SWAP gate, denoted as CSWAP,  has  the form
\eq
\label{cswap}
\textrm{CSWAP}(\theta)=\left(\begin{array}{cccc}
               1 & 0 & 0 & 0 \\
               0 & \cos{\frac \theta 2} & -i\sin{\frac \theta 2} & 0 \\
               0 & -i\sin{\frac \theta 2}& \cos{\frac \theta 2} & 0 \\
               0 & 0 & 0 & 1 \\
             \end{array}
           \right),
\en
where the nontrivial two-dimensional submatrix is a single-qubit gate $R_x(\theta)$. The control unitary gate $CU$ is defined as $CU=|0\rangle_L\langle 0|\otimes I_2+|1\rangle_L\langle 1| \otimes U$ with a single-qubit gate $U$. When the gate $U$ is the Pauli gate $X$, the $CU$ gate is the well known CNOT gate.

About the quantum control construction of two-qubit gates, we start with a four-level system for simplicity. The free Hamiltonian $\hat{H}_0$ and the interaction
Hamiltonian $\hat{H}_1$ are respectively given by
\eq \hat{H}_0=\sum_{n=0}^3 \tilde{E}_0(n) \widetilde{|n\rangle}_L\widetilde{\langle n|}, \quad \hat{H}_1=\hbar\sum_{n,n'=0}^3 \tilde{a}_{nn'}
 \widetilde{|n\rangle}_L\widetilde{\langle n'|}
\en
with  energy eigenvalues  $\tilde{E}_0(0) < \tilde{E}_0(1)< \tilde{E}_0(2) < \tilde{E}_0(3)$, where the basis states in a logical two-qubit Hilbert space are
respectively denoted as $\widetilde{|0\rangle}_L=|0\rangle_L|0\rangle_L$, $\widetilde{|1\rangle}_L=|0\rangle_L|1\rangle_L$, $\widetilde{|2\rangle}_L=|1\rangle_L|0\rangle_L$
and $\widetilde{|3\rangle}_L=|1\rangle_L|1\rangle_L$. The effective Hamiltonian $\hat{H}^I_-$ (\ref{low hamiltonian})  is
\eqa\label{four_level_general_form}
\hat{H}^I_{-}=\frac{\hbar\epsilon}{2}\left(
    \begin{array}{cccc}
    0 & \tilde{a}_{01}e^{i\varphi}e^{-i(\tilde{\omega}_{10}-\omega_d)t} & \tilde{a}_{02}e^{i\varphi}e^{-i(\tilde{\omega}_{20}-\omega_d)t} & \tilde{a}_{03}e^{i\varphi}e^{-i(\tilde{\omega}_{30}-\omega_d)t} \\
    \tilde{a}_{10}e^{-i\varphi}e^{i(\tilde{\omega}_{10}-\omega_d)t} & 0 & \tilde{a}_{12}e^{i\varphi} e^{-i(\tilde{\omega}_{21}-\omega_d)t}& \tilde{a}_{13}e^{i\varphi}e^{-i(\tilde{\omega}_{31}-\omega_d)t} \\
    \tilde{a}_{20}e^{-i\varphi}e^{i(\tilde{\omega}_{20}-\omega_d)t} & \tilde{a}_{21}e^{-i\varphi}e^{i(\tilde{\omega_{21}}-\omega_d)t} & 0 & \tilde{a}_{23}e^{i\varphi}e^{-i(\tilde{\omega}_{32}-\omega_d)t} \\
    \tilde{a}_{30}e^{-i\varphi} e^{i(\tilde{\omega}_{30}-\omega_d)t}& \tilde{a}_{31}e^{-i\varphi} e^{i(\tilde{\omega}_{31}-\omega_d)t}& \tilde{a}_{32}e^{-i\varphi}e^{i(\tilde{\omega}_{32}-\omega_d)t} & 0  \\
    \end{array}
    \right). \nonumber
 \ena

When the driving frequency satisfies $\omega_d=\tilde{\omega}_{21}$, the Rabi oscillation Hamiltonian $\hat{H}^I_{Rabi}$ for the on-resonant transition is
\eq
\hat{H}^I_{Rabi}=\frac{\hbar\epsilon}{2}\left(
    \begin{array}{cccc}
    0 & 0 & 0 & 0 \\
    0 & 0 & \tilde{a}_{12}e^{i\varphi} & 0 \\
    0 & \tilde{a}_{21}e^{-i\varphi} & 0 & 0 \\
    0 & 0 & 0 & 0  \\
    \end{array}
    \right).
\en
Denote $\tilde{a}_{21}=|\tilde{a}_{21}|e^{-i\tilde{\varphi}_{21}}$ and $\tilde{\varphi}_{21}^\prime=\tilde{\varphi}_{21}+\varphi$ and $\hat{n}_{21}'=(\cos\tilde{\varphi}'_{21},-\sin\tilde{\varphi}_{21}',0)$.  At $\tilde{\varphi}_{21}^\prime=0$, the unital vector
$\vec{n}_{21}^\prime$ becomes the axis $\vec{e}_x$, then the unitary evolution operator
(\ref{rabi evolution_operator}) has the form of the CSWAP gate (\ref{cswap}).

For the sake of the CNOT gate,  the Rabi oscillation Hamiltonian in the four-level system at $\omega_d=\tilde{\omega}_{32}$ is
\eq
\hat{H}^I_{Rabi}=|1\rangle_L\langle 1| \otimes \frac{\hbar\epsilon}{2}\left(
    \begin{array}{cc}
     0 & \tilde{a}_{23}e^{i\varphi}  \\
    \tilde{a}_{32}e^{-i\varphi} & 0  \\
    \end{array}
    \right).
\en
Set $\tilde{a}_{32}=|\tilde{a}_{32}|e^{-i\tilde{\varphi}_{32}}$ and $\tilde{\varphi}_{32}^\prime=\tilde{\varphi}_{32}+\varphi$ and $\hat{n}_{32}'=(\cos\tilde{\varphi}'_{32},-\sin\tilde{\varphi}_{32}',0)$. The unitary evolution operator
(\ref{rabi evolution_operator}) is just the $CU$ gate with the single-qubit gate  $U=R_{n'_{32}}(\epsilon |\tilde{a}_{32}|t)$.
At $\tilde{\varphi}'_{32}=0$ and $t=\frac {\pi} {\epsilon |\tilde{a}_{32}|}$, the gate $U$ is the Pauli gate $-i X$, and
the $CU$ gate is denoted as $\textrm{CNOT}'$. Under the local action of the phase gate $S$, the CNOT gate is obtained by
$\textrm{CNOT}=(S\otimes I_2)\textrm{CNOT}'$.

\subsection{Further research on  quantum control of integrable systems}

About further research, first, it is worthwhile investigating the application of the quantum KAM theorem to the control of integrable systems for universal
quantum computation. Here is an interdisciplinary research on the quantum KAM theorem, the rotating-wave approximation and the quantum control. Second,
concerning the construction of two-qubit gates, it is important to  present a rigourous and complete study on
the frequency selection of a four-level system from an $N$-level quantum system. Third, when the control parameter $\epsilon$ and the amplitudes $\tilde{a}_{nn'}$ are time-dependent, the optimal control  \cite{LM14} can be introduced for the construction of quantum gates.

\section{Quantum computation in central force fields}

We study universal quantum computation in central potential fields by resonant quantum control. We discuss various encodings of a logical qubit in terms of
action variables of integrable systems, together with the constructions of logical single-qubit gates. Then we think about how to build up the continuous
SWAP gates. Finally, we make remarks on further research.

\subsection{Representation of a qubit via action variables }

The Hamiltonian for a set of identical free integrable systems is given by $\hat{H}_0=\sum_{i=0}^L \hat{H}_0^{(i)}$ where the $\hat{H}_0^{(i)}$ is the Hamiltonian of
the $i$th subsystem and the $L$ is the number of subsystems. The energy eigenvector of the entire system is just a tensor product of each subsystem's energy eigenvectors. For a one-dimensional (or two-dimensional or three-dimensional) central potential field, its action variable eigenstate is respectively denoted by $|n\rangle^{(i)}=|n_r\rangle^{(i)}$ (or $|n\rangle^{(i)}=|n_r, n_\theta\rangle^{(i)}$ or  $|n\rangle^{(i)}=|n_r, n_\theta, n_\varphi \rangle^{(i)}$). The control Hamiltonian $\hat{H}_1$ has the same form as (\ref{control hamiltonian}) with the driving frequency $\omega_d$.

\subsubsection{Qubit encoded in a single action variable}

Consider the representation of a logical qubit in a single integrable system. Take the $i$th subsystem. Denote the basis of the qubit Hilbert space as
$|0\rangle^{( i)}_L=|n\rangle^{(i)}$ and $|1\rangle^{( i)}_L=|n'\rangle^{(i)}$. With the radical action variable in a three-dimensional central potential field,
the logical qubit basis states are denoted by  $|0\rangle^{(i)}_L=|n_r, n_\theta, n_\varphi \rangle^{(i)}$ and
 $|1\rangle^{(i)}_L=|n'_r, n_\theta, n_\varphi \rangle^{(i)}$. For example, the ground state $|0\rangle^{(i)}_L=|000 \rangle^{(i)}$
and the first excited state $|1\rangle^{(i)}_L=|100 \rangle^{(i)}$.

For non-degenerate integrable models such as anharmonic oscillators and the Coulomb potential with perturbation term, we construct single-qubit logical gates
by frequency selection with  $\omega_d=\omega_{10}\neq 0$, where $\hbar \omega_{10}=E^{(i)}_0(n')-E^{(i)}_0(n)$. But for degenerate models
including harmonic oscillators and Coulomb potentials, the selective driving at $\omega_d=\omega_{10}$ induces on-resonant transitions
beyond the subspace of the logical qubit, so that the construction of quantum gates via the resonance method fails.

With the angular action variable such as $j_\theta$, the logical qubit in a two-dimensional non-degenerate integrable model is represented by
$|0\rangle^{(i)}_L=|n_r, n_\theta\rangle^{(i)}$ and $|1\rangle^{(i)}_L=|n_r, n'_\theta \rangle^{(i)}$.  But in three-dimensional integrable models,
there is obviously a degeneracy introduced by the angular action variables $j_\theta$ and $j_\varphi$, such that the logical qubit
by $|0\rangle^{(i)}_L=|n_r, n_\theta, n_\varphi \rangle^{(i)}$ and $|1\rangle^{(i)}_L=|n_r, n'_\theta, n_\varphi \rangle^{(i)}$ is not allowed
by the resonance method.

\subsubsection{Qubit encoded in two or three action variables}

Look at the encoding of a logical qubit in terms of two action variables, such as the radical and angular action variables, or two angular action variables.
The frequency selection at $\omega_{10}\neq 0$ is applied to a logical qubit with the basis $|0\rangle^{(i)}_L=|n_r, n_\theta+1\rangle^{(i)}$
and $|1\rangle^{(i)}_L=|n_r+1, n_\theta \rangle^{(i)}$ in a two-dimensional non-degenerate integrable system and also for a qubit by
 $|0\rangle^{(i)}_L=|n_r, n_\theta\rangle^{(i)}$ and $|1\rangle^{(i)}_L=|n_r+1, n_\theta+1 \rangle^{(i)}$ or by
$|0\rangle^{(i)}_L=|n_r, n_\theta, n_\varphi \rangle^{(i)}$ and $|1\rangle^{(i)}_L=|n_r, n_\theta+1, n_\varphi+1 \rangle^{(i)}$ in both degenerate and non-degenerate
systems. But the resonant quantum control at $\omega_{10}=0$ instead of the frequency selection is needed for a logical qubit by
$|0\rangle^{(i)}_L=|n_r, n_\theta+1\rangle^{(i)}$ and $|1\rangle^{(i)}_L=|n_r+1, n_\theta \rangle^{(i)}$ or by
$|0\rangle^{(i)}_L=|n_r, n_\theta, n_\varphi+1 \rangle^{(i)}$ and $|1\rangle^{(i)}_L=|n_r, n_\theta+1, n_\varphi \rangle^{(i)}$ in a two-dimensional degenerate system.

 For a logical qubit encoded in terms of three action variables: $|0\rangle^{(i)}_L=|n_r, n_\theta, n_\varphi\rangle^{(i)}$ and $|1\rangle^{(i)}_L=|n'_r, n'_\theta, n'_\varphi \rangle^{(i)}$, for example, $|0\rangle^{(i)}_L=|000\rangle^{(i)}$ and $|1\rangle^{(i)}_L=|111\rangle^{(i)}$, we still apply the selective driving at $\omega_{10}\neq 0$ for
 the construction of single-qubit gates.

\subsubsection{Qubit encoded in two or three integrable systems}

Take a logical qubit represented by action variables of two independent identical integrable systems with the basis
$|0\rangle^{( i,i+1)}_L=|n\rangle^{(i)} |m\rangle^{(i+1)}$ and $|1\rangle^{( i,i+1)}_L=|n'\rangle^{(i)}|m'\rangle^{(i+1)}$.
For non-degenerate integrable
systems, we select a logical qubit by $|0\rangle^{( i,i+1)}_L=|n_r,n_\theta\rangle^{(i)} |n_r,n_\theta\rangle^{(i+1)}$ and
$|1\rangle^{( i,i+1)}_L=|n_r+1,n_\theta\rangle^{(i)} |n_r+1,n_\theta\rangle^{(i+1)}$ and apply the selective driving at $\omega_{10}\neq 0$ for the construction of
single-qubit gates; we also present a logical qubit by $|0\rangle^{( i,i+1)}_L=|n_r,n_\theta\rangle^{(i)} |n_r+1,n_\theta\rangle^{(i+1)}$ and
$|1\rangle^{( i,i+1)}_L=|n_r+1,n_\theta\rangle^{(i)} |n_r,n_\theta\rangle^{(i+1)}$ with the resonant control at $\omega_{10}=0$.
The latter case is the  dual-rail representation of a logical qubit in optical photon quantum computation \cite{NC2011}.

For a logical qubit encoded in three independent identical integrable systems, we choose the basis states as
$|0\rangle^{( i,i+1)}_L=|n\rangle^{(i)} |m\rangle^{(i+1)}  |p\rangle^{(i+2)}$ and $|1\rangle^{( i,i+1)}_L=|n'\rangle^{(i)}|m'\rangle^{(i+1)} |p'\rangle^{(i+2)}$.
For example, with three identical one-dimenisonal integrable systems, there is a natural logical qubit,
$|0\rangle^{( i,i+1)}_L=|0\rangle^{(i)} |0\rangle^{(i+1)}  |0\rangle^{(i+2)}$ and $|1\rangle^{( i,i+1)}_L=|1\rangle^{(i)}|1\rangle^{(i+1)} |1\rangle^{(i+2)}$, which
is a known quantum error-correction code in fault-tolerant quantum computation \cite{NC2011}.

\subsection{The construction of continuous SWAP gates}

Consider a conventional encoding of two qubits into two independent identical systems.
An orthonormal basis of the Hilbert space of the $i$th  and $i+1$th logical qubits is
\eqa
|00\rangle_L^{(i,i+1)}=|0\rangle^{(i)}_L |0\rangle^{(i+1)}_L, \quad |01\rangle_L^{(i,i+1)}=|0\rangle^{(i)}_L |1\rangle^{(i+1)}_L, \nonumber\\
|10\rangle_L^{(i,i+1)}=|1\rangle^{(i)}_L |0\rangle^{(i+1)}_L, \quad |11\rangle_L^{(i,i+1)}=|1\rangle^{(i)}_L |1\rangle^{(i+1)}_L.
\ena
These basis states are  rewritten as
\eq
 \widetilde{|0\rangle}_L=|00\rangle_L^{(i,i+1)},\,\, \widetilde{|1\rangle}_L=|01\rangle_L^{(i,i+1)},\,\,
 \widetilde{|2\rangle}_L=|10\rangle_L^{(i,i+1)},\,\, \widetilde{|3\rangle}_L=|11\rangle_L^{(i,i+1)},
\en
with the energy eigenvalues respectively given by
\eq
\tilde{E}_0(0)=E_0(00),\,\, \tilde{E}_0(1)=E_0(01),\,\, \tilde{E}_0(2)=E_0(10),\,\,  \tilde{E}_0(3)=E_0(11).
\en
Hence the frequency for a quantum jump between two basis states is denoted as
$\tilde{\omega}_{ij}=(\tilde{E}_0(i) - \tilde{E}_0(j))/\hbar$ with $i, j=0, 1, 2, 3$.

For simplicity, we concentrate on logical qubits in one-dimensional systems (or  encoded only by radical action variables).
The basis of the Hilbert space of the $i$th logical qubit is $|0\rangle^{(i)}_L=|n_r\rangle^{(i)}$ and
$|1\rangle^{(i)}_L=|n_r+1\rangle^{(i)}$, while the $i+1$th qubit is $|0\rangle^{(i+1)}_L=|n_r\rangle^{(i+1)}$ and
$|1\rangle^{(i+1)}_L=|n_r+1\rangle^{(i+1)}$. Obviously, $\tilde{\omega}_{10}=\tilde{\omega}_{32}$ and
 $\tilde{\omega}_{20}=\tilde{\omega}_{31}$, so that these four frequencies can not be exploited for the construction of
 quantum gates via frequency selection.

 There are several ways of setting up continuous SWAP gates (\ref{cswap}). First, when $\tilde{\omega}_{21}=0$, the method of the construction of
 a single-qubit gate at $\omega_d=0$ can be applied to build up the $R_x(\theta)$ gate in the subspace spanned by $\widetilde{|1\rangle}_L$
 and $\widetilde{|2\rangle}_L$. Second, when $\tilde{\omega}_{30} \neq 0$, with
 the driving frequency $\omega_d=\tilde{\omega}_{30}$, the frequency selection is applied for the construction of the $R_x(\theta)$ gate
 in the subspace spanned by $\widetilde{|0\rangle}_L$ and $\widetilde{|3\rangle}_L$.
 Third, if $|1\rangle^{(i)}_L=|n_r+2\rangle^{(i)}$, then $\tilde{\omega}_{21}\neq 0$. So  the
 resonant control with the driving frequency $\omega_d=\tilde{\omega}_{21}$ allows one to construct
 the $R_x(\theta)$ gate in the subspace spanned by $\widetilde{|1\rangle}_L$ and $\widetilde{|2\rangle}_L$.

\subsection{Remarks on further research in central force fields}

About further research of quantum computation in central potential fields, we focus on its physical realization in various experiments as well as
its fault-tolerance via the quantum KAM theorem. As above, we study universal quantum computation in central force fields by the resonance control,
so that we explore its experimental realization under the guidance of the well known  approaches to quantum computation \cite{NC2011},
which also use the resonance driving to construct quantum gates. In addition, we investigate quantum computation using large action variables,
such as quantum computation with the Rydberg states of the hydrogen atom,  from the viewpoint of the Bohr correspondence principle \cite{Born27}.

\section{The Birkhoff normal form for  universal quantum computation}

Besides the resonant quantum control, a general method of constructing two-qubit gates \cite{NC2011} is to study the unitary time evolution operator
determined by  the interaction between two single qubits. For example, the NMR quantum computation directly uses the spin-spin interaction, and the
optical photon quantum computation realizes an indirect interaction of two photons via non-linear Kerr media. So in the generalized harmonic oscillator
quantum computation, one can introduce interaction terms between harmonic oscillators for the purpose of building up two-qubit gates.

The extended harmonic oscillator quantum computation has a natural interpretation based on the Birkhoff normal form \cite{Arnold99} at
a neighborhood of an equilibrium point of a Hamiltonian system. Roughly speaking, the Hamiltonian $H_0$ in the Birkhoff normal form contains
a polynomial of simple harmonic oscillator Hamiltonians together with other relevant terms. When the Hamiltonian $H_0$ is integrable, it is a Birkhoff
integrable system near the equilibrium. For example,  when a polynomial of simple harmonic oscillator Hamiltonians is the Hamiltonian $H_0$, it gives
rise to a Birkhoff integrable system.

A classical Birkhoff normal form describes small oscillations of a Hamiltonian system around its equilibrium point. Suppose the classical Hamiltonian
as a sum of kinetic terms $p_i^2/2 m_i$ and a potential term $V(q_i)$ with particle mass $m_i$, coordinate $q_i$ and momentum $p_i$ for $i$th particle.
Assume $q_i=0$ as its equilibrium point with $p_i=0$ and $\partial V/\partial q_i|_{q_i=0}=0$. Hence the approximate Hamiltonian at the neighborhood of
the equilibrium point is derived by the Taylor expansion,
\eq
\label{classical Birkhoff}
H_0=\sum_{i=1}^n \left( {\frac {p_i^2} {2 m_i}} + \frac 1 2 m_i \omega^2_i q_i^2 \right) + \cdots,
\en
with classical frequencies $\omega_i$ defined by $ m_i\omega_i^2 =\partial^2 V  /\partial q_i^2|q_i=0$, where the term in the bracket is just a summation
of simple harmonic oscillator Hamiltonians. In equation (\ref{classical Birkhoff}), we assume a set of incommensurate frequencies $\omega_i$ up to order $k$,
satisfying the equation $\sum_{i=1}^n k_i \omega_i \neq 0$ for integer $k$, with $1\le \sum_{i=1}^n |k_i| \le k$. Then there exists a series of canonical
transformations leading to a Hamiltonian, which is a Birkhoff normal form of order $k$. When the order $k$ is an arbitrary integer,
the frequencies $\omega_1,  \omega_2, \cdots, \omega_n$ are incommensurate in a general sense, so that the resulting Birkhoff normal form represents
a Birkhoff integrable system.

A quantized Birkhoff normal form is obtained from canonical quantization of a classical Birkhoff normal form.
The quantum Birkhoff normal form of order $k$ is a sum of a polynomial of simple harmonic oscillator Hamiltonians $\hat{\tau}_i$ up to integer
order $[k/2]$ with other terms of higher order.  For example, the Hamiltonian $\hat{H}_0$ up to order 4 for universal quantum computation
has the form
\eq
\label{Birkhoff order 4}
 \hat{H}_0 =\sum_{i} c_i \hat{\tau}_i + \frac 1 2 \sum_{i,j} c_{ij} \hat{\tau}_i \hat{\tau}_j + \cdots,
\en
with linear coefficients $c_i$ and coupling coefficients $c_{ij}$, where the dots stand for higher order terms in $\hat{\tau}_i$. Here a two-qubit gate
can be constructed either by the unitary time evolution of coupling terms, or by frequency selection in resonant quantum control because the driving frequencies
are incommensurate due to the contribution of anharmonic oscillator terms.

To obtain a Birkhoff normal form in terms of action-angle variables, we either apply a canonical transformation from canonical coordinates and momenta to action-angle
variables, or directly calculate the Taylor expansion of an integrable system Hamiltonian $H_0(J)$ at the neighborhood of an invariant torus specified by a given value
of $J_i$ as $J_i|_o$ or $J|_o$. For example, we consider the Taylor expansion given by
\eq
\label{LM taylor expansion}
H_0(J)=H_0(J|_o)+\sum_i \omega^{cl}_i \Delta J_i+\frac 1 2 \sum_{i,j=1}^n \frac{\partial^2 H_0}{\partial J_i\partial J_j}|_{J|_o} \Delta J_i\Delta J_j+\cdots,
\en
with the classical frequency $\omega^{cl}_i=\partial {H_0}/{\partial J_i}|_{J|_o}$ and  small parameter $\Delta J_i=J_i-J_i|_o$. Via canonical quantization, the $\Delta J_i$
is replaced by the operator $\Delta\hat{J}_i$, which can be interpreted as a simple harmonic oscillator Hamiltonian with frequency $\omega^{cl}_i$, so that
$\Delta \hat{J}_i \Delta \hat{J}_j$ represents the coupling between harmonic oscillator Hamiltonians. As a matter of fact, the approximation model denoted by (\ref{LM taylor expansion}) plays a crucial role in Lloyd and Montangero's proposal on quantum computation in integrable systems \cite{LM14}.

Note that a Birkhoff normal form \cite{Arnold99} is an approximation of a Hamiltonian system at the neighborhood of an equilibrium point (or an invariant torus
or a periodic motion). It can be described either in terms of simple harmonic oscillator Hamiltonian variables or  directly using action variables.
The relationship between such two descriptions has to be made clear for sake of quantum computation using action variables. Moreover, a quantum Birkhoff
normal form can be either defined as a quantization of a classical Birkhoff normal form or as an approximation of a quantum Hamiltonian at the neighborhood of
an equilibrium point. Obviously, two resulting quantum models are not the same, so that we are rather careful in the study of quantum computation
via the Birkhoff normal form.

About physical realizations of quantum computation \cite{NC2011} via the Birkhoff normal form, first, we expect that it presents a mathematical framework for superconductor quantum computation using anharmonic oscillator potential. Second, when a quanta of a harmonic oscillator is a photon, we view optical photon quantum computation as
an example for quantum computation via the Birkhoff normal form. Third, when a quanta of a harmonic oscillator is a phonon in condensed matter physics, we refer to the ion trap quantum computation for a possible interesting study on the Birkhoff normal form.  Furthermore, regarding fault-tolerant quantum computation using the Birkhoff normal form,
we choose the total Hamiltonian in the quantum KAM theorem as the Birkhoff normal form of order 4, see (\ref{Birkhoff order 4}), which is a sum of linear terms, coupling terms
and perturbation terms.

\section{Discussion: quantum computation in integrable systems}

In the research article \cite{LM14},  Lloyd and Montangero have recently made a clear statement that a quantum
integrable system with a global control field is capable of performing universal quantum computation. They choose quantum action
variables of integrable systems to represent qubits, and explain how to apply the resonant driving theory to the construction of
universal quantum gates.

The key point is to observe that the first-order approximation of the Hamiltonian of a classical non-degenerate integrable system is
a collection of uncoupled harmonic oscillators with incommensurate frequencies. According to \cite{NC2011}, quantized harmonic oscillators
with control fields are essential in known quantum computation models, such as cavity quantum electrodynamics model, nuclear magnetic
resonance model and ion trap model. Therefore, canonical quantization of the first-order approximation of integrable systems can be used
to perform universal quantum computation.

Besides the construction of a universal quantum gate set, Lloyd and Montangero have investigated other interesting research topics
\cite{LM14}: the effectiveness of the rotating-wave approximation has an interpretation in the KAM theorem; the complexity of quantum
computation in integrable system has been analyzed; the optimal control is introduced to deal with a case that the coupling between the control
and the system is not constant; a strongly chaotic quantum system with a global control is not capable of performing universal quantum computation.

Figure 1 presents a sketch on quantum computation in integrable models.  There are two approaches of constructing a quantum  approximation model.
The one is obtained as an approximation of a quantum integrable model, whereas the other is derived from canonical quantization of an
approximation of a classical integrable model. Generally, they lead to different quantum approximation models. Note that a path from quantum
integrable model to quantum computation via quantum control shows a regular procedure for performing quantum computation, and a path
from quantum approximation model to quantum computation describes a routine approach to quantum computation using harmonic oscillators.

\begin{figure}[!htp]
  \centering
    \includegraphics[width=11cm]{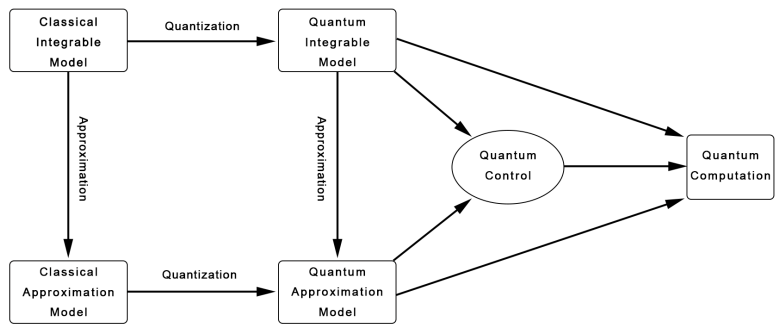}
  \caption{A sketch of quantum computation in integrable systems. There are  two essentially different ways of constructing a quantum approximation model.
  }
  \label{quantum_computation_1}

\end{figure}

As a matter of fact, how to perform quantum computation in integrable systems has been a long-term research project investigated by few groups in mathematical physics.
Integrable quantum computation \cite{Zhang11} is defined as quantum computation associated with the integrable condition. For example, the integrability condition is
the quantum Yang--Baxter equation, so that a multi-particle factorized scattering is identified with a quantum circuit model, where quantum computation without quantum
control is considered rather seriously. Here, Figure 1 describes the previous research \cite{Zhang11} by a path from quantum
integrable model to quantum computation with (or without) quantum control.

About further research of quantum computation in integrable systems, we wish to make a unified description for both quantum computation using the integrability condition and
quantum computation via the quantum KAM theorem. For example, we reformulate quantum computation using action variables as a special example for integrable quantum computation.
That is, we suggest that ``integrable quantum computation'' is a good notion for all possible quantum computations in integrable systems.¡¡¡¡¡¡¡¡ ¡¡

\section{Concluding remarks}\label{conclusion}

Motivated by Lloyd and Montangero's research proposal \cite{LM14} on quantum computation in integrable systems, we investigate quantum computation using action variables in the paper. We understand quantum computation using action variables as fault-tolerant quantum computation associated with the quantum KAM theorem \cite{RL87}.  For the Liouville integrable systems in central potential fields, we calculate quantum action variables by the quantum Hamilton--Jacobi equation \cite{LP83} and then construct quantum gates by the method of resonant quantum control. For the Birkhoff normal forms \cite{Arnold99} defining the Birkhoff integrable systems, we make a brief discussion on a generalized quantum computation using harmonic oscillators.

After all, quantum computation using action variables is a worthwhile research subject concerning quantum computation, quantum mechanics and mathematical physics. It sheds a light on further research of various topics including fault-tolerant quantum computation, the quantum KAM theorem and the quantum normal forms of Hamiltonian systems. For example, there is an interdisciplinary research among the quantum Hamilton--Jacobi equation, the quantum KAM theorem and the quantum Birkhoff normal forms. First of all, three of them emerge so naturally in quantum computation using action variables. Second, all of them are originally investigated in the study of non-linear differential equations. Third, a series of quantum canonical transformations are used in both the quantum KAM theorem and the Birkhoff normal form.

\section*{Acknowledgements}

 Yong Zhang has been supported by the NSF of China (Grant No. 11574237) on Integrable Quantum Computation. He is greatly  indebted to Xiaoyuan Li,
 Mo-Lin Ge, Louis H. Kauffman, Yong-Shi Wu, and Lu Yu \cite{Zhang11} for encouraging him to perform a long-term research on  quantum computation in
 integrable systems. He thanks Jing Shi, Quanlin Jie and Yijian Du for kind support during the research, thanks Lei Jing, Ye Wang, Yihua Liu and Yuping An
 for an initial study on topics of the paper, and thanks all participants in the course of Quantum Information and Quantum Computation in Wuhan University
 from 2012 to 2021.


\end{document}